\documentclass[11pt]{elsart}
\usepackage[dvips]{graphicx}

\usepackage{array}
\usepackage{amssymb}
\usepackage{amsmath}
\usepackage{hhline}
\usepackage{longtable}
\usepackage{dcolumn}
\usepackage{bm}
\usepackage{subfigure}

\begin{document}

\begin{frontmatter}

\title{A Synthetical Weights' Dynamic Mechanism for Weighted Networks}

\author[ad1,ad2]{Lujun Fang}
\ead{fanglujun@fudan.edu.cn}
\author[ad1,ad2]{Zhongzhi Zhang}
\ead{zhangzz@fudan.edu.cn}
\author[ad1,ad2]{Shuigeng Zhou\corauthref{zsz}}
\corauth[zsz]{Corresponding author.}
\ead{sgzhou@fudan.edu.cn}
\author[ad3]{Jihong Guan}
\ead{jhguan@mail.tongji.edu.cn}
\address[ad1]{Department of Computer Science and Engineering, Fudan
University,\\ Shanghai 200433, China}%
\address[ad2]{Shanghai Key Lab of Intelligent Information Processing, Fudan
University,\\ Shanghai 200433, China}%
\address[ad3]{Department of Computer Science and Technology, Tongji University,
\\4800 Cao'an Road, Shanghai 201804, China}%

\begin{abstract}
We propose a synthetical weights' dynamic mechanism for weighted
networks which takes into account the influences of strengths of
nodes, weights of links and incoming new vertices. Strength/Weight
preferential strategies are used in these weights' dynamic
mechanisms, which depict the evolving strategies of many
real-world networks. We give insight analysis to the synthetical
weights' dynamic mechanism and study how individual weights'
dynamic strategies interact and cooperate with each other in the
networks' evolving process. Power-law distributions of strength,
degree and weight, nontrivial strength-degree correlation,
clustering coefficients and assortativeness are found in the model
with tunable parameters representing each model. Several
homogenous functionalities of these independent weights' dynamic
strategy are generalized and their synergy are studied.


\end{abstract}

\date{}
\end{frontmatter}

\section{Introduction}

Complex networks~\cite{AlBa02,DoMe02,Ne03,BoLaMoChHw06,BoSaVe07}
depict a great many real-world networks like the scientific
collaboration networks
(SCN)~\cite{Ne01a,Newman01,BaJeNeRaScVi02,LiWuWaZhDiFa07}, World
Wide Web (WWW)~\cite{AlJeBa99}, world-wide airport networks
(WAN)~\cite{BaBaPaVe04,LiCa04} and so on. Simple binary
networks~\cite{AlBa02,DoMe02,Ne03} are used to depict the
topological aspects of these real-world networks. Degree
distribution and degree related clustering coefficients can be
analyzed from the model. Typically, Barab\'asi and Albert proposed
a linear preferential attachment model on which most other binary
models are based on (BA model\cite{BaAl99}). However, real-world
networks often contains far more information than that binary
networks can express, since relations in the networks are not
necessarily binary. Therefore links with weights are introduced to
emphasize the importance of heterogenous relations between nodes
in networks. Barrat, Barth\'elemy and Vespignani first build a
model for weighted networks based on preferential attachment
mechanism (BBV model~\cite{BaBaVe04a,BaBaVe04b}). Since then
various models have been derived to mimic diverse real-world
networks~\cite{BoLaMoChHw06,BoSaVe07,WaWaHuYaQu05,WuXuWa05,GoKaKi05,MuMa06,XiWaWa07,WaHuZhWaXi05,WaHuWaYa06},
all of which emphasize a different kind of weights' dynamic
mechanism.

The flourishing research on various weights' dynamics mechanisms
roots in the the heterogenous behaviors of real-life networks.
Current research has already covered most part of typical
behaviors of weights' dynamics. However, the interaction and
operations of these scattered weights' dynamics are poorly studied
and their underlying homogeneity are still not know. In our paper,
we generalize weights' dynamics with monotonous weights' growth
and focus on interactions and cooperations among them.

With a close scrutiny into prevailing weights' dynamic models
depicting real-world networks we find that there are three main
sources of weights' increment dynamics: the variation of traffic
caused by introducing of new
nodes~\cite{BoLaMoChHw06,BoSaVe07,WaWaHuYaQu05,WuXuWa05,GoKaKi05,MuMa06},
the increment of links' weights based on links' weights
themselves~\cite{LeCh07}, and the increment of weights based on
strength of two ends of the
link~\cite{XiWaWa07,WaHuZhWaXi05,WaHuWaYa06}. Most weights'
increment dynamics can be grouped into these three sources. Many
works have done to empirically validate the classification, among
them Newman gives the most comprehensive and convincing
experiments~\cite{Ne01sta}. In this paper we will give numerical
and analytical study to the synthetical weights' dynamic model
which comprises the three above-mentioned mechanisms, and reveal
the scale-free characteristics and nontrivial clustering
coefficients and assortativeness of the model.

Our paper is organized as follows. In Sec. $2$ we define basic
definitions and terms to represent the weighted network and gives
a brief review of related works done. In Sec. $3$ we depict our
model and give clear definition of three weights' dynamic
mechanisms. In Sec. $4$ we analytically calculate the mathematical
expression for probability distributions of strength and weight,
degree-strength correlations and related attributes. In Sec. $5$
we preform simulations to mimic the proposed mechanisms and
analyze the experiment results in detail. In Sec. $6$ we give a
conclusion to the paper.

\section{Definitions}

Weighted networks can be represented by a adjacent matrix $W$
where $w_{ij}$ defines the weight of link between vertices $i$ and
$j$. $w_{ij} = 0$ indicates that there is no link between vertices
$i$ and $j$. Therefore topological and weighted information can
both be revealed from $W$. Matrix $W$ is symmetrical therefore
$w_{ij} = w_{ji}$. Degree $k_i$ defines the number of vertices
vertex $i$ is linked with, and strength $s_i$ defines the total
weights of links that ends in the particular vertex $i$. $s_i$ can
be written as $s_i = \sum_{j} w_{ij}$, and $k_i$ can be written as
$k_i = \sum_{j} sgn(w_{ij})$, where $sgn()$ is the signum
function. $P(s)$, $P(k)$, $P(w)$ defines the probability
distribution of strength, degree and weight. Previous studies have
revealed that in many weighted networks, $P(s)$, as well as $P(k)$
and $P(w)$, displays a power-law distribution as $P(s) =
s^{-\gamma_{s}}$, $P(k) = k^{-\gamma_{k}}$ and $P(w) =
w^{-\gamma_{w}}$. There is also a power-law correlation between
$s$ and $k$ that $s = k^{\alpha}$. Clustering coefficients and
assortativeness of weighted network is also studied and the
details will be discussed later.

\section{The Model}

The model proposed in this paper starts from an initial
configuration of $N_{0}$ vertices fully connected by links with
weight $w_{0}=1$ ($N_{0}$-clique). At each time step, the network
evolves under two coupled mechanisms: topological growth and
weights' dynamics. Weights' dynamics are discussed in detail in
this paper, where all three sources of weights' dynamics are taken
into account.

\subsection{Topological Growth}

At each time step, a new vertex $n$ is introduced into the network
and connected to $p$ existing vertices $i$. Vertices are chosen
according to the strength preferential probability
\begin{equation}
\Pi_{n\rightarrow i}=\frac{s_{i}}{\sum_{j}s_{j}} \,,
\end{equation}
and the weight of this new link is set to $w_{0}=1$.

\subsection{Weights' Dynamics}

There are three sources of weights' dynamics: the local increment
of weights triggered by the introduction of the new vertex, the
self-increment of weights based on the weight of each link, and
the mutual selection dynamics focusing on creation and
reinforcement of links between existing vertices based on their
strengths. These three weights' dynamics mechanisms interact and
cooperate during the evolution of the network. There are several
suggestive independent works for these three sources weights'
dynamics: Barrat, Barth\'elemy, Vespignani first suggest the local
rearrangement model considering the impact of incoming vertices.
Dorogovtsev, Mendes's work Wen-Xu Wang's works suggest the mutual
selection model which well depict the second source. ********'s
work initializes the idea of weights' self-increment although the
idea is not comprehensively studied yet. There are lots of works
done that empirically proving the validity of the three sources.
Newman in his work by empirically studying the scientific
collaboration network suggests that two scientists would have
better chance to enforce their collaboration if they already have
a lot of works together, and scientists are more likely to develop
new collaborative relationships if they already have relatively
large numbers of collaborators~\cite{Ne01sta}. Collaborative
relationships of scientists and their collaborators are analogous
to weights and degree in a network. Degree can be generalized to
strength if we take into account the amount of collaborations
between each pairs of collaborators in stead of a binary
expression.

The introduction of new vertex brings variation in traffic across
the network. For simplicity, we restrict the variation to the
neighborhood of vertex $i$ which has just been chosen to link with
the new vertex. An overall increment of $\delta$ is introduction
at each time step. The increment is distributed among the
neighborhood of $\Gamma (i)$ according to weight preferential
mechanism:
\begin{equation}
w_{ij}\rightarrow w_{ij}+ \delta \frac{w_{ij}}{s_{i}} \,.
\end{equation}
The strengths of $i$ and all $j \in \Gamma (i)$ are also increased
as a result of the increment of weights in the neighborhood of
$i$. Considering the probability of vertices $i$ been chosen, the
increment of $w_{ij}$ can be rewritten as
\begin{align}
\Delta w_{ij} &= p\frac{s_{i}}{\sum_{k}s_{k}}\delta
\frac{w_{ij}}{s_{i}} + p\frac{s_{j}}{\sum_{k}s_{k}}\delta
\frac{w_{ij}}{s_{j}} \nonumber\\
&= 2p\delta\frac{w_{ij}}{\sum_{k}s_{k}} \,.
\end{align}

At each time step, $n$ existing links are chosen to increase
according to the weight preferential probability:
\begin{equation}
w_{ij}\rightarrow w_{ij}+ n \frac{w_{ij}}{\sum_{k,l}w_{kl}}
\end{equation}
Each chosen link is increased by $w_0 = 1$. The links with larger
weights always have more chance to reinforcement.

At each time step, each existing vertex $i$ selects $m$ vertices
according to the strength preferential mechanism:
\begin{equation}
\Pi_{i\rightarrow j}=m\frac{s_{j}}{\sum_{k}s_{k}-s_{i}} \,.
\end{equation}
There would be a alteration in links between $i$ and $j$ if and
only if $i$ and $j$ have mutually selected each other. The
probability that the linking condition between $i$ and $j$ changes
can be defined to be:
\begin{align}
\Pi_{i,j}&= m\frac{s_{j}}{\sum_{k}s_{k}-s_{i}}
m\frac{s_{i}}{\sum_{k}s_{k}-s_{j}} \nonumber\\
&=
m^{2}\frac{s_{i}s_{j}}{(\sum_{k}s_{k}-s_{i})(\sum_{k}s_{k}-s_{j})}\nonumber\\
&\approx m^{2}\frac{s_{i}s_{j}}{(\sum_{k}s_{k})^{2}}\,.
\end{align}
If there is not a link between $i$ and $j$,
a new link with assigned weight $w_0 = 1$ will be added. If there
is already a link between $i$ and $j$, the link will be increased
by $w_0 = 1$.

These three weights' dynamics mechanisms interact and cooperate
during the the process of network development. Synthesize all the
these three mechanisms, the increment of weights can be
represented to be
\begin{equation}
w_{ij}\rightarrow w_{ij} + 2p\delta\frac{w_{ij}}{\sum_{k}s_{k}} +
n \frac{w_{ij}}{\sum_{k,l}w_{kl}} +
m^{2}\frac{s_{i}s_{j}}{(\sum_{k}s_{k})^{2}} \,.
\end{equation}
Noticing the fact that $\sum_{k,l}w_{kl} = \frac{1}{2}\sum_{k}
s_{k}$, we can rewrite the above equation as
\begin{equation}
w_{ij}\rightarrow w_{ij} + (2p\delta+2n)
\frac{w_{ij}}{\sum_{k}s_{k}} +
m^{2}\frac{s_{i}s_{j}}{(\sum_{k}s_{k})^{2}} \,.
\end{equation}

\section{Evolution and Distribution of Degree, Strength and
Weight}

\begin{figure}[t]
  \centering\includegraphics[width=10cm]{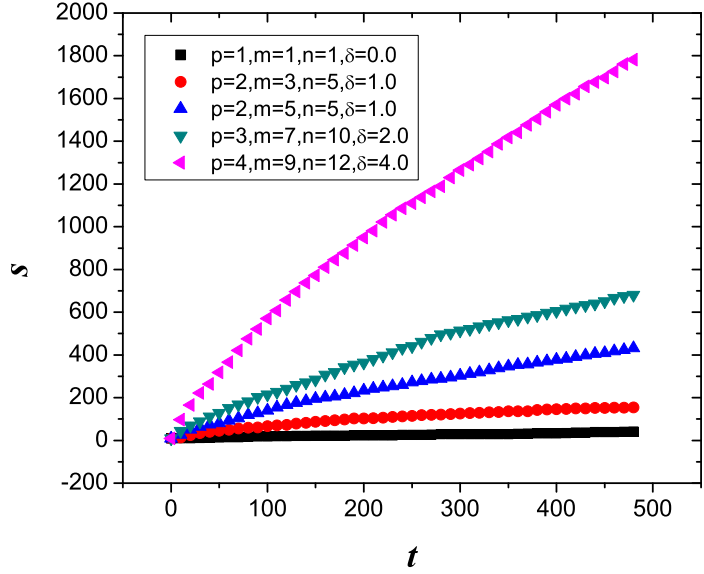}
  \caption{Evolution of degree $s_i$ with time for vertex $i=1$, the time span is from 1 to 5000 and the result is average of 10 individual experiments. Various values for $p$, $m$, $n$, $\delta$ are chosen to display the interplay of different parameters.} \label{figSevo}
\end{figure}

Using the continuous approximation, we can assume that $s$, $k$,
$w$, $t$ are all continuous. Therefore we get
\begin{equation}\label{dwdt}
\frac{dw_{ij}}{dt} = (2p\delta+2n) \frac{w_{ij}}{\sum_{k}s_{k}} +
m^{2}\frac{s_{i}s_{j}}{(\sum_{k}s_{k})^{2}} \,.
\end{equation}
There are two sources contributing the increment of strength
$s_{i}$, one is the weights' dynamic and the other is linking with
the new added node. Therefore the increment of $s_{i}$ can be
written as
\begin{align}\label{dsdt0}
\frac{ds_{i}}{dt} &= \sum_{j}\frac{dw_{ij}}{dt} +
p\frac{s_{i}}{\sum_{k}s_{k}} \nonumber\\
&= (m^2+2p\delta+2n+p)\frac{s_{i}}{\sum_{k}s_{k}} \,.
\end{align}
The sum of strength of all nodes $\sum_{k}s_{k}$ at time $t$ can
be calculated as
\begin{align}\label{sum_si}
\sum_{k}s_{k}(t) & =
 \sum_{1}^{t} s_{i} \nonumber\\ &=
\int_{0}^{t}\sum_{k}\frac{ds_{k}}{dt}+pt\nonumber\\
&=(m^2+2p\delta+2n+2p)t \,,
\end{align}
and using this equation we can rewrite Eq. (\ref{dsdt0}) as
\begin{equation}
\frac{ds_{i}}{dt}=\frac{m^2+2p\delta+2n+p}{m^2+2p\delta+2n+2p}\frac{s_{i}}{t}
\,.
\end{equation}
With the initial condition $s_{i}(t=i) = 1$, we can integrate the
above equation to obtain
\begin{equation}\label{si}
s_{i}(t) = \left( \frac{t}{i}
\right)^{\frac{m^2+2p\delta+2n+p}{m^2+2p\delta+2n+2p}}
\end{equation}
From the equation we can see that three parameters $m$, $\delta$
and $n$ cooperatively and interactively govern the growing speed
of strength $s_{i}$. Is is really amazing to find out that all
three sources of weights' dynamics influence the growing speed of
strength in similar ways. The simulation of evolution of $s_{i}$
is given in Fig.~\ref{figSevo}. We see how $m$, $\delta$, $n$, $p$
contribute to the evolution of $s_i$ independently by fixing three
other parameters. We also show how these four parameters interact
by varying them the same time as indicated by the above equation.
We see $s_i$ display a power-law distribution as $t$ evolves, and
variable $m$ contribute larger alteration in $s_i$ with relatively
small amount of increment.

\begin{figure}[t]
  \centering\includegraphics[width=10cm]{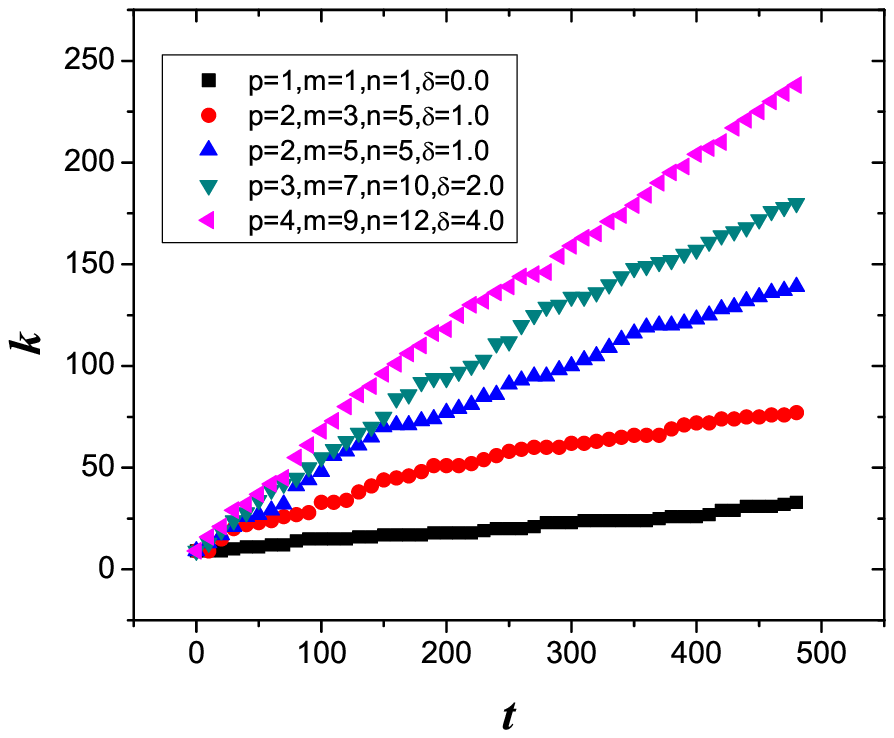}
  \caption{Evolution of degree $k_i$ with time for vertex $i=1$, the time span is from 1 to 5000 and the result is average of 10 individual experiments. Various values for $p$, $m$, $n$, $\delta$ are chosen to display the interplay of different parameters.} \label{figKevo}
\end{figure}

\begin{figure}[t]
  \centering\includegraphics[width=10cm]{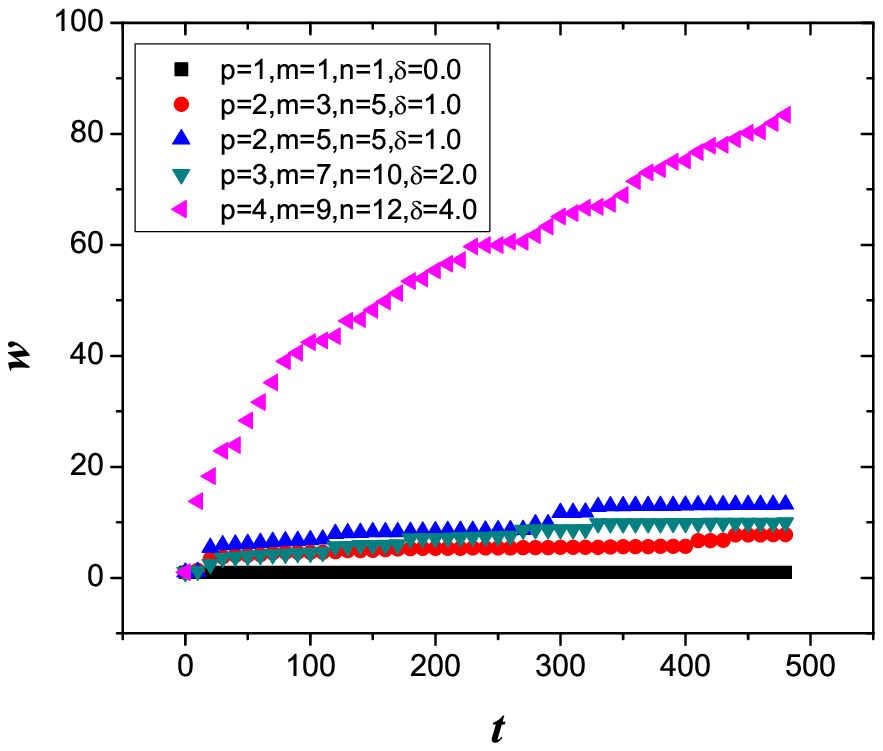}
  \caption{Evolution of degree $w_i$ with time for vertex $i=1$ and $j=2$, the time span is from 1 to 5000 and the result is average of 10 individual experiments. Various values for $p$, $m$, $n$, $\delta$ are chosen to display the interplay of different parameters.} \label{figWevo}
\end{figure}

\begin{figure}[t]
  \centering\includegraphics[width=10cm]{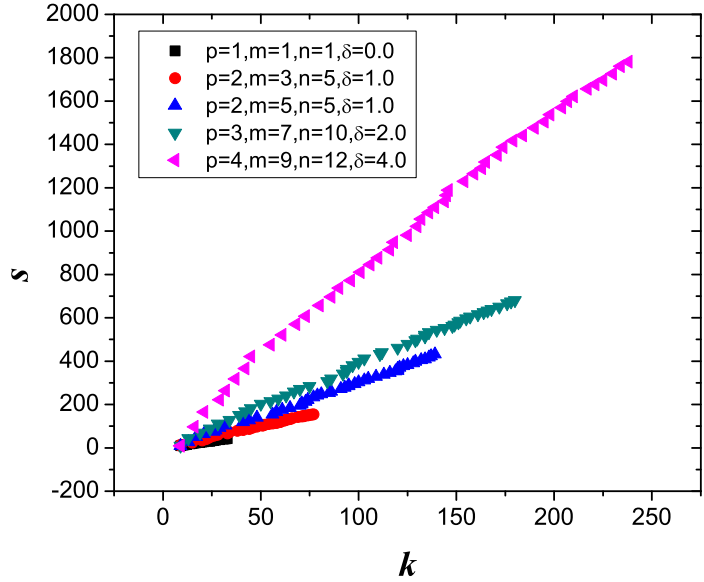}
  \caption{Nontrivial correlation between degree $k$ and strength $s$, the size of the network is 5000 and the result is average of 10 individual experiments. Various values for $p$, $m$, $n$, $\delta$ are chosen to display the interplay of different parameters.} \label{figSKevo}
\end{figure}

\begin{figure}[t]
  \centering\includegraphics[width=10cm]{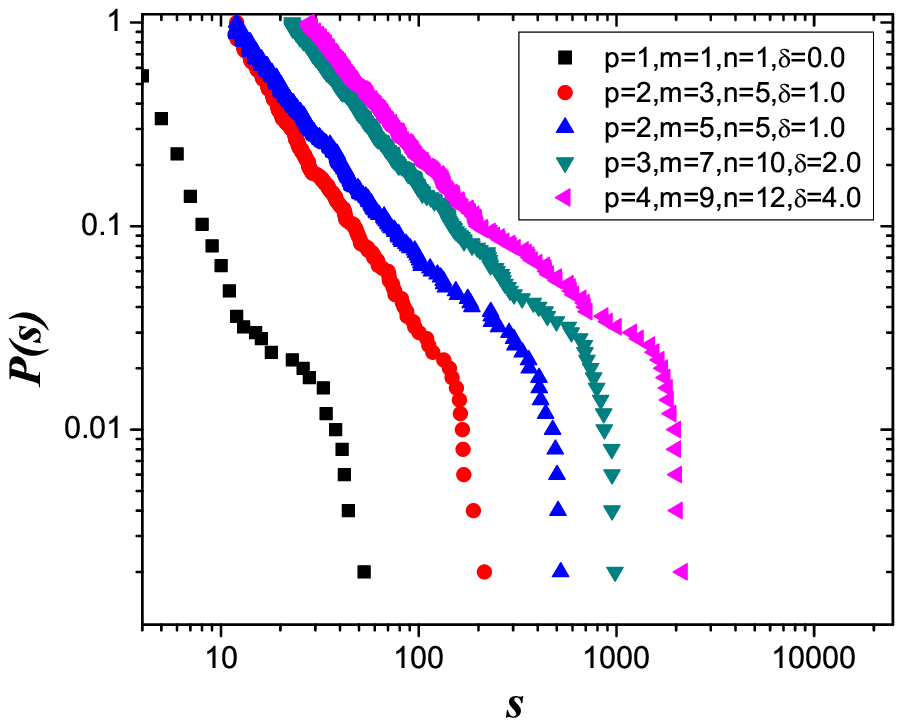}
  \caption{Power-law probability distribution of strength $s$, the size of the network is 5000 and the result is average of 10 individual experiments. Various values for $p$, $m$, $n$, $\delta$ are chosen to display the interplay of different parameters.} \label{figSdis}
\end{figure}

\begin{figure}[t]
  \centering\includegraphics[width=10cm]{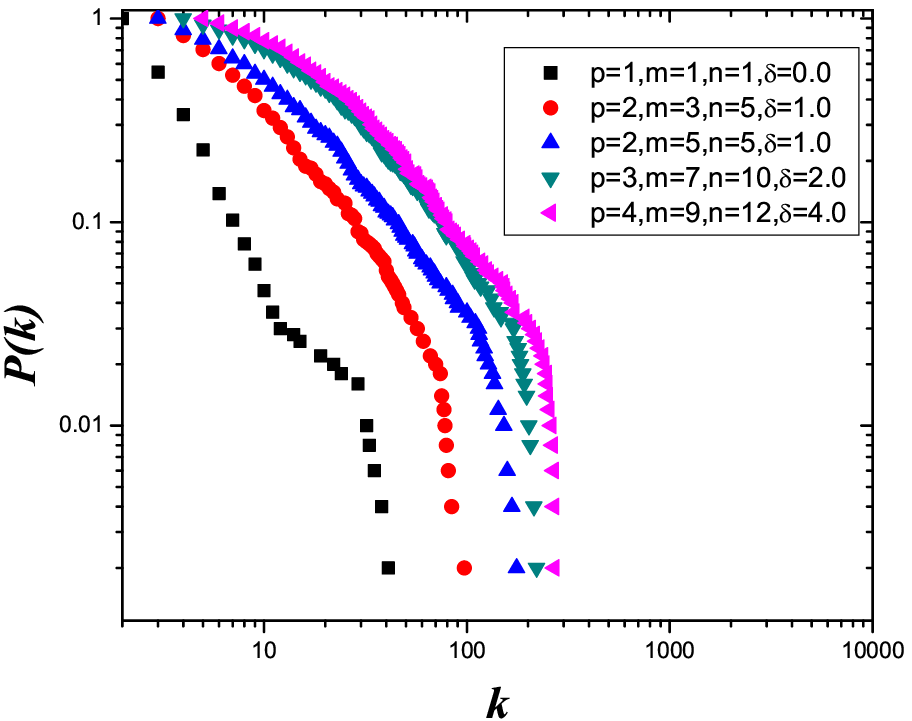}
  \caption{Power-law probability distribution of degree $k$, the size of the network is 5000 and the result is average of 10 individual experiments. Various values for $p$, $m$, $n$, $\delta$ are chosen to display the interplay of different parameters.} \label{figKdis}
\end{figure}

\begin{figure}[t]
  \centering\includegraphics[width=10cm]{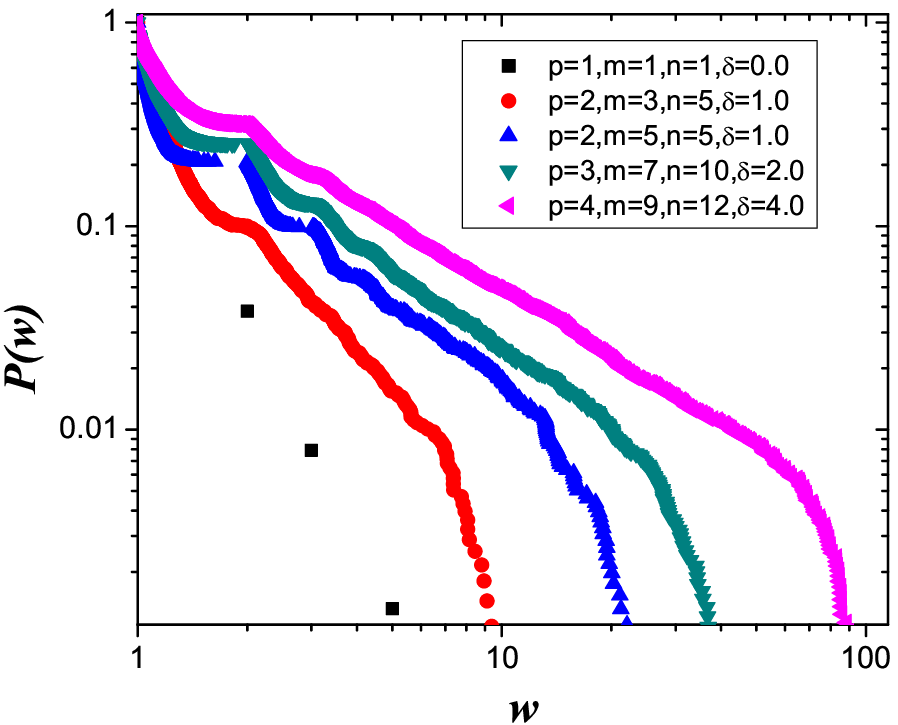}
  \caption{Power-law probability distribution of weight $w$, the size of the network is 5000 and the result is average of 10 individual experiments. Various values for $p$, $m$, $n$, $\delta$ are chosen to display the interplay of different parameters.} \label{figWdis}
\end{figure}

The knowledge of the time evolution of the various quantities
allows us to compute their statistical properties. The incoming
time $t_{i}$ of each vertex $i$ is uniformly distributed in
$[0,t]$ and the strength probability distribution can be written
as
\begin{equation}
P(s,t) = \frac{1}{t+N_{0}}\int_{0}^{t}\delta(s-s_{i}(t))dt_{i} \,,
\end{equation}
where $\delta(x)$ is the Dirac delta function. Using Eq.
(\ref{si}) we can get in the infinite size limit $t\rightarrow
\infty$ the distribution $P(s) \sim s^{-\gamma_s}$ with
\begin{equation}
\gamma_s=2+\frac{p}{m^2+2p\delta+2n+p} \,.
\end{equation}
The degree probability distribution $P(k)\sim k^{-\gamma_k}$ can
be obtained by combining $s \sim k^\beta$ with Eq. (\ref{si}).
From the equation of  the conservation of probability
\begin{equation}
\int_{0}^{\infty}P(k)dk=\int_{0}^{\infty}P(s)ds
\end{equation}
we can get
\begin{equation}
P(k)\sim P(s)\frac{ds}{dk} \sim s^{-\gamma_s}\beta k^{\beta-1}
\sim \beta k^{-[\beta(\gamma_s-1)+1]} \,,.
\end{equation}
Therefore we get $\gamma_k=\beta(\gamma_s-1)+1$ in $P(k)\sim
k^{-\gamma_k}$. The simulations of $k_{i}$ and $P(k)$¡¡are given
in Fig.~\ref{figKevo} and Fig.Fig.~\ref{figKdis} with resemble the
figures for $s_{i}$ and $k_{i}$. The power-law correlation between
$s$ and $k$ is reveal by Fig.~\ref{figSKevo}, where we fix $m=p=1$
and tune $\delta$ and $n$ as we need. .

The evolution and distribution of weight can be calculated
similarly as we deal with strength. Combine Eq. (\ref{dwdt}), Eq.
(\ref{sum_si}) ,Eq. (\ref{si}), and define $a=m^2+2p\delta+2n+2p$,
we get
\begin{equation}
\frac{dw_{ij}}{dt}=\frac{2p\delta+2n}{a}\frac{w_{ij}}{t}+
\frac{m^2}{a^2}\left(\frac{1}{ij}\right)^{1-\frac{1}{a}}
t^{-\frac{2}{a}}
\end{equation}
we can integrate the above equation and get
\begin{equation}
w_{ij} \sim t^{\frac{m^2+2p\delta+2n}{m^2+2p\delta+2n+2p}}
\end{equation}
for large $t$. Therefore $P(w)$ can be represented as
$P(w)=w^{-\gamma_{w}}$ with
$\gamma_{w}=2+\frac{2p}{m^2+2\delta+2n}$. The simulations of
$w_{ij}$ and $P(w)$ are given representatives in
Fig.~\ref{figWevo} and Fig.~\ref{figWdis}.

\section{Clustering Coefficients and Assortativeness}

\begin{figure}[t]
  \centering\includegraphics[width=10cm]{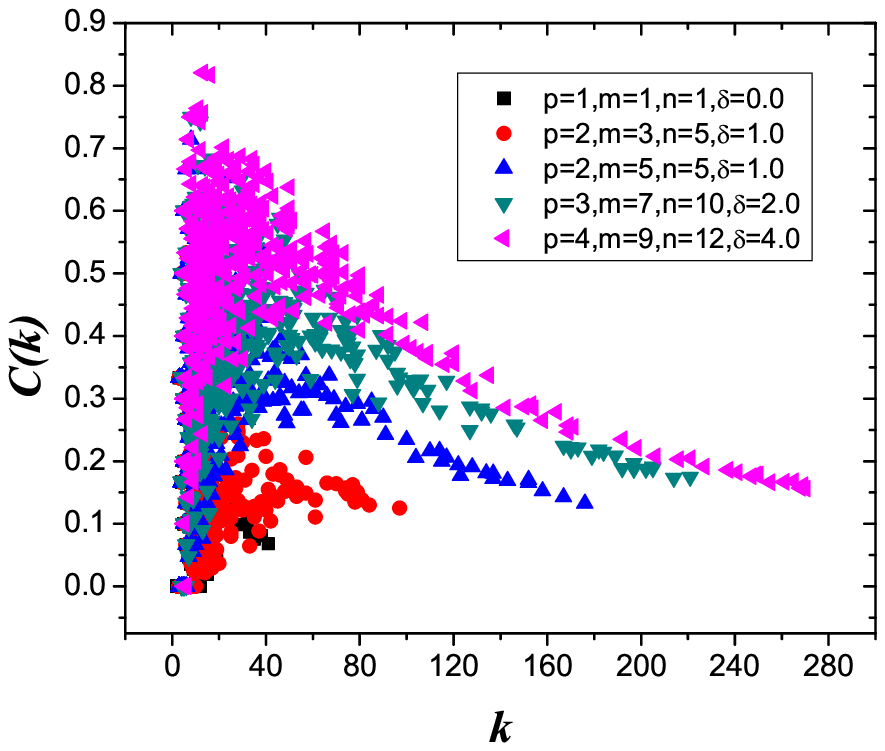}
  \centering\includegraphics[width=10cm]{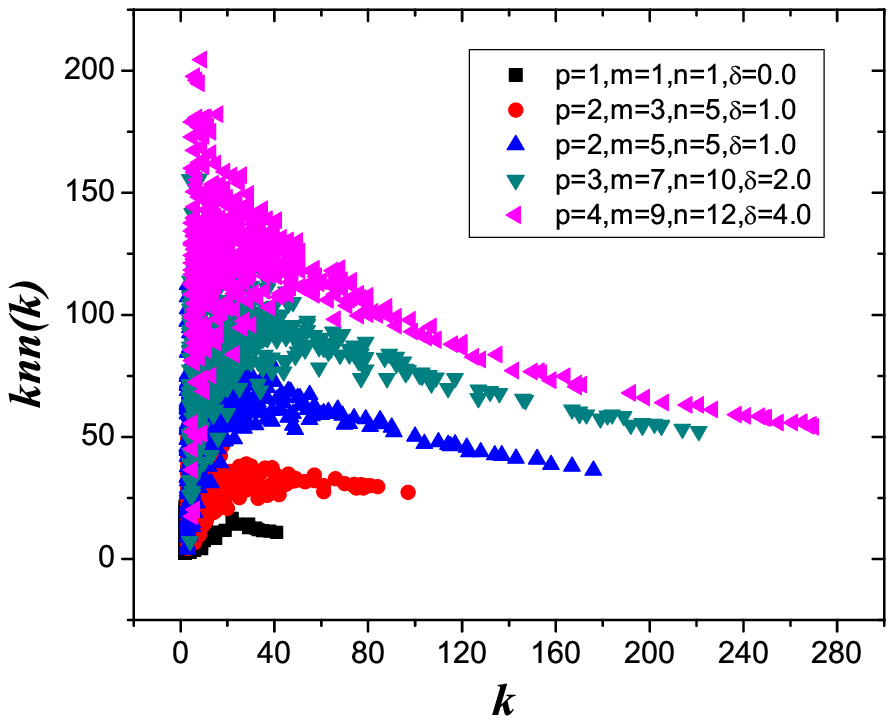}
  \caption{(a) Clustering coefficients for different degree $k$.
   (b) Average nearest-neighbor degree for different degree $k$. The size of the network is 5000 and the result is average of 10 individual experiments. Various values for $p$, $m$, $n$, $\delta$ are chosen to display the interplay of different parameters.} \label{figClustering}
\end{figure}

Clustering coefficients depict connectivity among neighborhood of
given vertices. The local clustering coefficient $c_i$ for
specific vertex $i$ is defined to be $c_i =
\frac{1}{C_{k_{i}}^{2}}\sum_{j<k}sgn(w_{ij}w_{jk}w_{ik})$, which
denotes the percentage of number of triangles in the neighborhood
of $i$ to the number of potential triangles there. $c_{i}$ reveals
how close vertices in the neighborhood of $i$ are related. The
average $c_i$ of all vertices is denoted as $C$, and the average
of $c_i$ for vertices with degree $k$ is denoted as $C(k)$.
Average degree of nearest neighbor $knn,i$ for vertex $i$ is also
studied, as well as $knn(k)$ for the average of $knn,i$ of
vertices with degree of $k$. $knn$ reveals the assortativeness of
a network.

We perform numerical experiment the simulate the growth of network
and analyze the clustering coefficient and assortativeness under
the synthetical weights' dynamic mechanism. In
Fig.~\ref{figClustering}(a) we show how clustering coefficients
vary with tunable variables. The scale-free attributes $C$
measures the overall density of triangles in the network. We can
see that $C$ is consistent with the network size $N$ while varying
positively when those tunable variables change. The evolution of
$knn$ is show in Fig.~\ref{figClustering}(b) which suggest the
tunable assortativeness of the network.

\section{Conclusion}

In summary, we propose a new model for weighted networks which
synthesizes three sources of weights' dynamic: local weights'
rearrangement raised by introduction of new vertex; self-increment
of weights according to weights' preferential strategy; weights'
creation and reinforcement proportional to strengths of both ends
of nodes. Three sources independently contribute to the evolution
and in the mean time also cooperatively interact. The homogenous
behaviors that these weights' dynamics display suggest there may
be some common underlying mechanisms that are not yet well
understood. This model would be a good start for a synthetical and
general understanding of weights' dynamic and hopefully our work
would be helpful for the future study.

\section*{Acknowledgements}
This research was supported by the National Natural Science
Foundation of China under Grant Nos. 60496327, 60573183, and
90612007. Zhongzhi Zhang also thank the support of the Postdoctoral
Science Foundation of China under Grant No. 20060400162, and Huawei
Foundation of Science and Technology.


\begin{thebibliography}{10}

\bibitem{AlBa02} R. Albert and A.-L. Barab\'asi,
       Rev. Mod. Phys. {\bf 74}, 47 (2002).

\bibitem{DoMe02} S. N. Dorogvtsev and J.F.F. Mendes,
Adv. Phys. {\bf 51}, 1079 (2002).

\bibitem{Ne03} M. E. J. Newman,
SIAM Review {\bf 45}, 167 (2003).

\bibitem{BoLaMoChHw06}
S Boccaletti, V Latora, Y Moreno, M. Chavezf, and D.-U. Hwanga,
Physics Report {\bf 424}, 175 (2006).

\bibitem{BoSaVe07}
K. Borner, S. Sanyal£¬and A. Vespignani£¬
  Ann. Rev. Infor. Sci. Tech. {\bf 41}, 537 (2007).

\bibitem{Ne01a}
M. E. J. Newman, Proc. Natl. Acad. Sci. U.S.A. {\bf 98} (2001)
404.

\bibitem{Newman01}
M. E.~J. Newman, Phys. Rev. E {\bf 64}, 016132 (2001).

\bibitem{BaJeNeRaScVi02}
A.-L. Barab\'asi, H. Jeong, Z. N\'eda. E. Ravasz, A. Schubert, and
T. Vicsek, Physica A {\bf 311}, 590 (2002).

\bibitem{LiWuWaZhDiFa07}
M Li, J Wu, D Wang, T Zhou, Z Di, Y Fan, Physica A {\bf 375}, 355
(2007).

\bibitem{AlJeBa99}
R. Albert, H. Jeong and  A.-L. Barab\'asi, Nature {\bf 401} (1999)
130.

\bibitem{BaBaPaVe04}
A. Barrat, M. Barth\'elemy, R. Pastor-Satorras, and A. Vespignani,
Proc. Natl. Acad. Sci. U.S.A. {\bf 101}, 3747 (2004).

\bibitem{LiCa04}
W. Li, and X. Cai, Phys. Rev. E {\bf 69}, 046106 (2004).

\bibitem{BaAl99} A.-L. Barab\'asi and R. Albert,
       Science {\bf 286}, 509 (1999).

\bibitem{BaBaVe04a} A. Barrat, M. Barth\'elemy, and A. Vespignani, Phys.
Rev. Lett. {\bf 92}, 228701 (2004).

\bibitem{BaBaVe04b} A. Barrat, M. Barth\'elemy, and A. Vespignani, Phys.
Rev. E {\bf 70}, 066149 (2004).

\bibitem{WaWaHuYaQu05} W.-X. Wang, B.-H. Wang, B. Hu, G. Yan, and Q. Ou, Phys. Rev.
Lett. \textbf{94}, 188702 (2005).

\bibitem{WuXuWa05} Z.-X. Wu, X.-J. Xu, and Y.-H. Wang, Phys. Rev. E {\bf 71}, 066124 (2005).

\bibitem{GoKaKi05} K.-I. Goh, B. Kahng, and D. Kim, Phys. Rev. E {\bf 72}, 017103 (2005).

\bibitem{MuMa06} G. Mukherjee and S. S. Manna, Phys. Rev. E {\bf 74}, 036111 (2006).

\bibitem{XiWaWa07} Y.-B. Xie, W.-X. Wang, and B.-H. Wang, Phys. Rev. E \textbf{75}, 026111 (2007).

\bibitem{WaHuZhWaXi05} W.-X. Wang, B. Hu, T. Zhou, B.-H. Wang,
Y.-B. Xie, Phys. Rev. E \textbf{72} 046140(2005).

\bibitem{WaHuWaYa06} W.-X. Wang, B. Hu, B.-H. Wang,
G. Yan, Phys. Rev. E \textbf{73} 016133(2005).

\bibitem{LeCh07}
C.C. Leung, H.F. Chau, Physica A {\bf 378}, 591-602 (2007).

\bibitem{Ne01sta}
M. E. J. Newman, Phys. Rev. E \textbf{64}, 025102 (2001).


%
\bibitem{FaFaFa99}
M. Faloutsos, P. Faloutsos and C. Faloutsos, Comput. Commun. Rev.
{\bf 29} (1999) 251.



\bibitem{JeToAlOlBa00}
H. Jeong, B. Tombor, R. Albert, Z.N. Oltvai and A.-L. Barab\'asi,
Nature {\bf 407} (2000) 651.

\bibitem{JeMaBaOl01}
H. Jeong, S. Mason, A.-L. Barab\'asi and Z. N. Oltvai, Nature {\bf
411} (2001) 41.




\bibitem{LiEdAmStAb01}
F. Liljeros, C.R. Edling, L. A. N. Amaral, H.E. Stanley, Y.
\AA{berg}, Nature {\bf 411} (2001) 907.




\bibitem{WaSt98} D. J. Watts and H. Strogatz,
        Nature (London) {\bf 393}, 440 (1998).




\bibitem{KrFrMaUlTa03}
A. E. Krause, K. A. Frank, D. M. Mason, R. E. Ulanowicz, and W. W.
Taylor, Nature (London) {\bf 426}, 282 (2003).


\bibitem{YoJeBaTu01}
S.H. Yook, H. Jeong, A.-L. Barab\'asi, Y. Tu, Phys. Rev. Lett.
{\bf 86}, 5835 (2001).

\bibitem{ZhTrZhHu03}
D. Zheng, S. Trimper, B. Zheng, P.M. Hui, Phys. Rev. E {\bf 67}
040102 (2003).



\bibitem{AnKr05}
T. Antal and P. L. Krapivsky, Phys. Rev. E {\bf 71} 026103 (2005).




\bibitem{BaRaVi01} A.-L. Barab\'asi, E. Ravasz, and T. Vicsek,
          Physica A  {\bf 299}, 559 (2001).

\bibitem{DoGoMe02} S. N. Dorogovtsev, A. V. Goltsev, and J. F. F. Mendes,
          Phys. Rev. E {\bf 65}, 066122 (2002).

\bibitem{CoFeRa04} F. Comellas, G. Fertin and A. Raspaud,
Phys. Rev. E {\bf 69}, 037104 (2004).

\bibitem{ZhRoZh07}
Z. Z. Zhang, L. L. Rong, and S. G. Zhou,
Physica A {\bf 377} (2007) 329.

\bibitem{JuKiKa02} S. Jung, S. Kim, and B. Kahng,
        Phys. Rev. E {\bf 65}, 056101 (2002).

\bibitem{RaSoMoOlBa02}
E. Ravasz, A.L. Somera, D.A. Mongru, Z.N. Oltvai, and A.-L.
Barab\'asi, Science {\bf 297}, 1551 (2002).

\bibitem{RaBa03}
E. Ravasz and A.-L. Barab\'asi, Phys. Rev. E {\bf 67}, 026112
(2003).

\bibitem{AnHeAnSi05} J. S. Andrade Jr., H. J. Herrmann, R. F. S. Andrade and L. R. da Silva,
Phys. Rev. Lett. {\bf 94}, 018702 (2005).

\bibitem{DoMa05} J. P. K. Doye and C. P. Massen,
Phys. Rev. E {\bf 71}, 016128 (2005).


\bibitem{ZhCoFeRo05}
Z. Z. Zhang, F. Comellas, G. Fertin and L. L. Rong,
 J. Phys. A {\bf 39}, 1811 (2006).


\bibitem{ZhRo05} Z. Z. Zhang, L. L. Rong, and Shuigeng Zhou,
 Phys. Rev. E, {\bf 74}, 046105
(2006).

\bibitem{Bobe05}E. Bollt, D. ben-Avraham, New Journal of Physics {\bf 7}, 26
(2005).

\bibitem{BeOs79}
A.N. Berker and S. Ostlund, J. Phys. C {\bf 12}, 4961 (1979).

\bibitem{HiBe06}
M. Hinczewski and A. N. Berker, Phys. Rev. E {\bf 73}, 066126
(2006).

\bibitem{CoOzPe00}
F. Comellas, J. Oz\'on, and J. G. Peters, Inf. Process. Lett.,
{\bf 76}, 83 (2000)

\bibitem{CoSa02}
F. Comellas and M. Sampels, Physica A {\bf 309}, 231 (2002).


\bibitem{ZhRoGo05}
Z. Z. Zhang, L. L. Rong and C. H. Guo, Physica A {\bf 363}, 567
(2006).


\bibitem{ZhRoCo05a} Z. Z. Zhang, L. L. Rong and F. Comellas,
J. Phys. A {\bf 39}, 3253 (2006).

\bibitem{DoMe05} S. N. Dorogvtsev and J.F.F. Mendes,
AIP Conf. Proc. {\bf 776}, 29 (2005).

\bibitem{Fr77} C.L. Freeman, Sociometry {\bf 40}, 35 (1977).


\bibitem{SzMiKe02}
G. Szab\'o, M. Alava, and J. Kert\'esz, Phys. Rev. E {\bf 66},
026101 (2002).


\bibitem{BoRi04}
B. Bollob\'as and O. Riordan, Phys. Rev. E {\bf 69}, 036114
(2004).


\bibitem{GhOhGoKaKi04}
C.-M. Ghima, E. Oh, K.-I. Goh, B. Kahng, and D. Kim, Eur. Phys. J.
B {\bf 38}, 193 (2004)


\bibitem{MsSn02}
S. Maslov and K. Sneppen,
Science {\bf 296}, 910 (2002).

\bibitem{PaVaVe01}
R. Pastor-Satorras, A. V\'azquez and A. Vespignani,
Phys. Rev. Lett. {\bf 87}, 258701 (2001).

\bibitem{VapaVe02}
A. V\'azquez, R. Pastor-Satorras and A. Vespignani,
Phys. Rev. E {\bf 65}, 066130 (2002).

\bibitem{Newman02}
M. E.~J. Newman,
Phys. Rev. Lett. {\bf 89}, 208701 (2002).

\bibitem{Newman03c}
M. E.~J. Newman,
Phys. Rev. E {\bf 67}, 026126 (2003).


\bibitem{ZhZh07}
Z. Z. Zhang and S. G. Zhou,
Physica A (in press), e-print cond-mat/0609270.  

\end{thebibliography}
\end{document}